\documentclass[journal,10pt]{IEEEtran}

\usepackage{graphicx}
\usepackage{enumerate}
\usepackage{cite}
\usepackage{bm}
\usepackage{color}
\usepackage{amsmath,amssymb,amsthm}
\usepackage{setspace}

\usepackage[linesnumbered,boxruled,vlined]{algorithm2e}
\usepackage[hidelinks]{hyperref}
\usepackage{config}

\usepackage[font=footnotesize,skip=1.8pt]{caption}
\begin{document}
\title{Malicious RIS versus Massive MIMO:\\Securing Multiple Access against RIS-based Jamming Attacks}
\author{Arthur Sousa de Sena, \textit{Member}, \textit{IEEE},
Jacek Kibi\l{}da, \textit{Senior Member}, \textit{IEEE},\\Nurul Huda Mahmood, 
André Gomes, \textit{Member}, \textit{IEEE}, Matti Latva-aho, \textit{Fellow}, \textit{IEEE}

\thanks{Arthur S. de Sena, Nurul H. Mahmood, and Matti Latva-aho are with the University of Oulu, Oulu, Finland (email: arthur.sena@oulu.fi, nurulhuda.mahmood@oulu.fi, matti.latva-aho@oulu.fi).}
\thanks{Jacek Kibi\l{}da and André Gomes are with the Commonwealth Cyber Initiative, Virginia Tech, USA (email: jkibilda@vt.edu, gomesa@vt.edu).}
\thanks{The research leading to this paper received support from the Smart Networks and Services Joint Undertaking (SNS JU) under the European Union’s Horizon Europe research and innovation programme within \href{https://hexa-x-ii.eu/}{Hexa-X-II project} (Grant Agreement No 101095759), the Academy of Finland under the \href{https://www.6gflagship.com/}{6G Flagship program} (Grant No 346208) and the academy project ReWIN-6G (Grant No 357120). This material is also based upon work supported by the National Science Foundation, under Grants No. 2326599 and 2318798, and the Commonwealth Cyber Initiative.}
\vspace{-2mm}}


\maketitle

\begin{abstract}
In this letter, we study an attack that leverages a reconfigurable intelligent surface (RIS) to induce harmful interference toward multiple users in massive multiple-input multiple-output (mMIMO) systems during the data transmission phase. We propose an efficient and flexible weighted-sum projected gradient-based algorithm for the attacker to optimize the RIS reflection coefficients without knowing legitimate user channels. To counter such a threat, we propose two reception strategies. Simulation results demonstrate that our malicious algorithm outperforms baseline strategies while offering adaptability for targeting specific users. At the same time, our results show that our mitigation strategies are effective even if only an imperfect estimate of the cascade RIS channel is available.
\end{abstract}

\begin{IEEEkeywords}
	Reconfigurable intelligent surface, massive MIMO, passive jamming, physical-layer security\vspace{-1mm}
\end{IEEEkeywords}

\IEEEpeerreviewmaketitle

\section{Introduction}

\ac{RIS} is a low-cost and low-power alternative to RF repeaters that is meant to enhance service availability and resilience of next-generation wireless networks by dynamically controlling the reflection of impinging electromagnetic waves. Coupled with a passive mode of operation, it is also an attractive technology in adversarial applications. A malicious actor may deploy a \ac{RIS} or hijack an existing one to easily trigger attacks that utilize impinging signals to attack wireless communication links.

Ample recent works have proposed a variety of attacks, including creating destructive multipath~\cite{lyu2020irs}, assisting an active jammer~\cite{wang2022wireless}, or an eavesdropper~\cite{chen2022malicious}. However, a potentially more effective way to attack legitimate systems with a \ac{RIS} hinges on exploiting vulnerabilities in wireless system design. For instance, active \ac{RIS} architectures can create amplified destructive millimeter-wave beamforming~\cite{Rev_Lin23}, or alter \ac{RIS} reflective coefficients between the transmission of data and control symbols, resulting in a substantial degradation in symbol error rate~\cite{alakoca2022metasurface}. Similarly, by introducing malicious reflections during the channel estimation process, a malicious \ac{RIS} can reduce the effectiveness of channel equalization~\cite{staat2022mirror}, physical layer key generation rate~\cite{li2022reconfigurable}, or corrupt the beamforming vectors ~\cite{yang2021novel}. Vulnerabilities in channel estimation are particularly critical for multiple access methods in \ac{mMIMO} like \ac{SDMA}. 

\ac{SDMA}-aided \ac{mMIMO} involves two stages: the \ac{CSI} acquisition, based on transmitted pilot sequences, and data transmission using a precoder design based on the acquired \ac{CSI}. The dependency on \ac{CSI} accuracy makes \ac{mMIMO} systems vulnerable to pilot contamination attacks triggered by active transmitters~\cite{wang2019pilot}. However, this attack can also be conducted without additional power by altering the \ac{RIS} phase shifts between pilot and data transmission. More worryingly, it was shown in~\cite{huang2023illegal,huang2023disco} that an attack against legitimate users can be made effective even with random \ac{RIS} phase shifts, named ``Disco Ball" attack.

In this letter, we propose a new and more potent version of the attack against downlink data transmission to multi-antenna users in a \ac{mMIMO} system employing \ac{SDMA}, whereby the malicious \ac{RIS} does not randomly select the phase shifts, as in the Disco Ball strategy~\cite{huang2023illegal,huang2023disco}, but efficiently optimizes them to increase the damage, all without the knowledge of legitimate channels. To accomplish this, we develop an efficient and flexible weighted-sum projected gradient-based algorithm to optimize the \ac{RIS} coefficients. Subsequently, we propose two reception strategies that require only an estimate of the effective cascade \ac{RIS} channel to mitigate the adverse effects of the attack. Simulation results demonstrate that our malicious algorithm outperforms baseline strategies in unleashing more powerful attacks and exhibits adaptability for targeting specific users. Our results also show that the proposed mitigation strategies are effective even when only an imperfect estimate of the cascade \ac{RIS} channel is available.
\vspace{.3mm}

\noindent  {\it Notation:}
The $i$th element of a vector $\mathbf{a}$ is denoted by $[\mathbf{a}]_i$, the $(ij)$ entry of a matrix $\mathbf{A}$ by $[\mathbf{A}]_{ij}$, the submatrix of $\mathbf{A}$ formed by its rows (columns) from $i$ to $j$ by $[\mathbf{A}]_{i:j,:}$ $\left([\mathbf{A}]_{:,i:j}\right)$. The transpose and Hermitian transpose of $\mathbf{A}$ are represented by $\mathbf{A}^T$ and $\mathbf{A}^H$, respectively, $\mathbf{I}_M$ is the $M\times M$ identity matrix, $\mathbf{0}_{M, N}$ is the $M\times N$ zero matrix, and $\diamond$ represents the Khatri-Rao product. The operator $\mathrm{vec}\{\cdot \}$ transforms an $M\times N$ matrix into a column vector, $\mathrm{vecd}\{\cdot \}$ converts the diagonal elements of an $M\times M$ matrix into a column vector, $\mathrm{diagm}\{\cdot \}$ creates an $M\times M$ diagonal matrix from the diagonal elements of an arbitrary $M\times M$ matrix, $\mathrm{diag}\{\cdot \}$ transforms a vector of length $M$ into an $M\times M$ diagonal matrix, $\angle(z)$ returns the phase of the complex number $z$, and $\mathrm{E}\{\cdot\}$ denotes expectation.

\section{System Model}

\begin{figure}
	\centering
	\includegraphics[width=.9\linewidth]{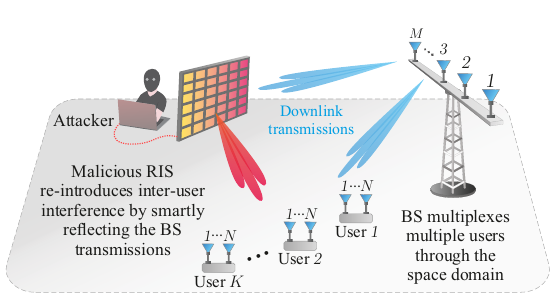}
	\caption{An attacker uses \iac{RIS} to perform an interference attack against data transmission in downlink multi-user \ac{mMIMO}.}\label{fig:sysmodel}
\end{figure}

We study a downlink multi-user \ac{mMIMO} system, consisting of one \ac{BS} employing $M$ antennas and $K$ spatially distributed users, represented by the index set $\mathcal{K} = \{1,2,\cdots, K\}$, where each user is equipped with $N$ antennas, such that $M\gg N$. The \ac{BS} spatially multiplexes the users using an \ac{SDMA} strategy, in which linear precoders are employed. In the presence of accurate \ac{CSI}, such precoders can efficiently tackle \ac{mMIMO} inter-user interference. However, as illustrated in Fig. \ref{fig:sysmodel}, the system is jammed by an attacker that controls a malicious \ac{RIS} comprising $L$ reflecting elements. It is assumed the attacker has enough computational power to acquire the \ac{CSI} of the cascade \ac{RIS} channel, but not the direct channels between the users and the BS. With this information in hand, the attacker can recycle impinging signals coming from the BS and passively re-introduce inter-user interference that conventional precoders cannot eliminate. Under the described scenario, the $k$th user receives the following signal
\begin{align}\label{rec_sig_01}
\mathbf{y}_{k} &= \big( \mathbf{F}_{k}^H
      \bm{\Theta}
    \mathbf{G} +  
    \mathbf{H}_{k}^H\big)  \sum_{i=1}^{K} \mathbf{P}_{i}\sqrt{P}\mathbf{x}_{i} + \mathbf{n}_{k}  \in \mathbb{C}^{N},
\end{align}
where $\mathbf{x}_{i} = [\sqrt{\alpha_{i,1}}x_{i,1}, \cdots, \sqrt{\alpha_{i,S}} x_{i,S}]^T\in \mathbb{C}^{S}$ is the data vector comprising $S$ symbols intended for the $i$th user, satisfying $\mathrm{E}\{|x_{i,s}|^2\} = 1$, with $\alpha_{i,s}$ denoting the power allocation coefficient for the $s$th symbol, $P$ represents the transmit power budget, $\mathbf{P}_{i} \in \mathbb{C}^{M \times S}$ is the precoding matrix responsible for multiplexing users in space, $\mathbf{n}_{k} \in \mathbb{C}^{N}$ is the noise vector\footnote{Since the attacker controls only a passive RIS with no active amplifiers, any thermal noise introduced by the RIS reflecting elements is negligible.}, whose entries follow the complex Gaussian distribution with zero mean and variance $\sigma^2$, and $\bm{\Theta}$ is the diagonal reflection matrix of the malicious \ac{RIS}, whose amplitude and phase-shift coefficients associated with the $l$th reflecting element satisfy the constraints $|[\bm{\Theta}]_{ll}| = 1$ and $\angle([\bm{\Theta}]_{ll}) \in [0, 2\pi], \forall l = 1, \cdots, L$. 
The matrices $\mathbf{H}_{k} \in \mathbb{C}^{M \times N}$, $\mathbf{G} \in \mathbb{C}^{L \times M}$, and $\mathbf{F}_{k} \in \mathbb{C}^{L \times N}$ represent the channels between the BS and the $k$th user (link BS-U), the BS and the RIS (link BS-RIS), and the RIS and the $k$th user (link RIS-U), respectively, and are assumed to have correlated entries.

\subsection{Precoding for User Multiplexing}\label{prec_sec}
To ensure the effectiveness of the passive \ac{RIS} jamming attack, the attacker maliciously stays idle during the \ac{BS} channel estimation phase. This means the \ac{BS} can rely only on the jamming-free channels observed on each BS-U link to construct the precoding matrix $\mathbf{P}_k$. The \ac{BS} then employs a two-layered spatial multiplexing strategy in which the desired precoder is structured as $\mathbf{P}_k \triangleq \mathbf{K}_k\mathbf{D}_k$, where $\mathbf{K}_k$ addresses inter-user interference and $\mathbf{D}_k$ maximizes the reception of intended signals at the $k$th user. By recalling the truncated eigendecomposition, the channel covariance matrix of the direct link BS-U for the $k$th user can be expressed as $\bm{\Sigma}_{k} \triangleq \mathrm{E}\{\mathbf{H}_{k} \mathbf{H}_{k}^H \} = \mathbf{\bar{U}}_{k} \bm{\bar{\Delta}}_{k} \mathbf{\bar{U}}^H_{k}$, where $\bm{\bar{\Delta}}_{k} \in \mathbb{R}^{r_k \times r_k}$ is a diagonal matrix comprising $r_k$ nonzero eigenvalues of $\bm{\Sigma}_{k}$, and $\mathbf{\bar{U}}_{k} \in \mathbb{C}^{M \times r_k}$ is a tall matrix containing the associated eigenvectors. The users' channel matrices can then be factorized as $\mathbf{H}_{k} = \mathbf{\bar{U}}_{k}\bm{\bar{\Delta}}_{k}^{\frac{1}{2}} \mathbf{\bar{H}}_{k}$, where $\mathbf{\Bar{H}}_{k} \in \mathbb{C}^{r_s \times N}$ is the reduced-dimension fast-fading channel matrix, whose entries follow the complex Gaussian distribution with zero mean and unity variance. Thus, the inter-user interference can be tackled if $ [\mathbf{\bar{U}}_{1}\hspace{1mm} \cdots \hspace{1mm}\mathbf{\bar{U}}_{k-1} \hspace{1mm}\mathbf{\bar{U}}_{k+1} \hspace{1mm} \cdots \hspace{1mm} \mathbf{\bar{U}}_{K}]^H \mathbf{K}_k = \bm{\Lambda}^H_k \mathbf{K}_k = \mathbf{0}, \forall k \in \mathcal{K}$, where $\bm{\Lambda}_k \in \mathbb{C}^{M \times \sum_{k'\neq k} r_{k'}}$ is a block matrix with the eigenvectors corresponding to the non-zero eigenvalues of non-intended users. This implies that $\mathbf{K}_k$ must be designed from the orthogonal complement of the column space of $\bm{\Lambda}_k$, which can be achieved with the aid of its full singular value decomposition (SVD). More specifically, let $\mathbf{U}_{\bm{\Lambda}_k}$ denote the left eigenvector matrix of $\bm{\Lambda}_k$ computed via SVD. Then, since the last $M - \sum_{k'\neq k} r_{k'}$ columns of $\mathbf{U}_{\bm{\Lambda}_k}$ provide the orthonormal basis for the desired orthogonal complement, the precoder for the $k$th user can be computed as
\begin{align}\label{BS_prec}
\mathbf{K}_k = \frac{1}{\sqrt{S}}\left[\mathbf{U}_{\bm{\Lambda}_k} \right]_{ :,\left(M - S + 1 \right) : M} \in \mathbb{C}^{M \times S},    
\end{align}
where the parameter $S$ sets the number of parallel symbols to be transmitted to each user, which must satisfy $S \leq \left( M - \sum_{k'\neq k} r_{k'} \right)$ and $S \leq \min\{r_{k}\}, \forall k \in \mathcal{K}$.

On the other hand, we design the precoder $\mathbf{D}_k$ to maximize the received power, i.e., $\underset{\mathbf{D}_k}{\max} \| \mathbf{H}^H_k \mathbf{K}_k \mathbf{D}_k \|^2_F$, at the $k$th user. This goal can be achieved through the full SVD of the projected channel $\mathbf{H}^H_k \mathbf{K}_k \in \mathbb{C}^{N\times S}$. Specifically, by recalling the SVD, we can decompose $\mathbf{H}^H_k \mathbf{K}_k =  \mathbf{\hat{U}}_{k} \bm{\hat{\Delta}}_{k} \mathbf{\hat{V}}^H_{k}$. Then, the inner precoder can be obtained as $\mathbf{D}_{k} = \mathbf{\hat{V}}_{k} \in \mathbb{C}^{S\times S}$.

\vspace{-1mm}
\section{RIS-Assisted Attack Design}
With the precoder in \eqref{BS_prec}, the \ac{BS} can effectively suppress inter-user interference experienced in the link BS-U. However, because the BS cannot detect the malicious RIS during channel estimation, inter-user interference propagating through the reflected BS-RIS-U link will inevitably be re-introduced to the users. Thus, the $k$th user will observe the following signal
\begin{align}\label{rec_sig_02}
\mathbf{y}_{k} = \mathbf{H}_{k}^H \mathbf{P}_{k}\sqrt{P}\mathbf{x}_{k} + \mathbf{F}_{k}^H
      \bm{\Theta}
    \mathbf{G}\sum_{i=1}^{K} \mathbf{P}_{i}\sqrt{P}\mathbf{x}_{i} + \mathbf{n}_{k}.
\end{align}

The attacker aims to re-introduce as much inter-user interference as possible in the system to degrade the detection of intended symbols at each targeted user $k \in \mathcal{K}$. To this end, the reflecting coefficients of the deployed RIS are tuned to steer all transmissions coming from the BS to jam multiple access.

The challenge is that the attacker is unlikely to have access to the legitimate BS-U channels or the BS precoders. Despite this, it can still exploit the RIS channels to boost inter-user interference in the system. In this case, the attacker will attempt to match the channels of the BS-RIS link with those of the RIS-U link by maximizing $\left\| \mathbf{F}_{k}^H \bm{\Theta}\mathbf{G}\right\|_F^2$. However, since a single RIS has been deployed, the attacker cannot find a global set of reflecting coefficients to achieve its objective optimally for all users. Alternatively, the attacker may apply a flexible jamming framework to tune the intensity of the attack for each user. This framework is implemented by computing the Pareto optimal points for the following weighted-sum problem:
\begin{subequations}\label{prob_1}
\begin{align}
    & \underset{\bm{\Theta}}{\max} 
     \sum_{k=1}^{K} \nu_{k} \left\| \mathbf{F}_{k}^H \bm{\Theta}\mathbf{G}
    \right\|_F^2,
    \label{prob_1a}\\[-1mm]
    &\text{s.t.} \hspace{2mm} \text{\small $|[\bm{\Theta}]_{mn}| = 1, \hspace{1mm} \forall m,n \in\{ 1,\cdots, L\} \hspace{1mm} | \hspace{1mm} m=n$},\label{prob_1b}\\[-1mm]
    & \hspace{6mm} \text{\small $|[\bm{\Theta}]_{mn}| = 0, \hspace{1mm} \forall m,n \in\{ 1,\cdots, L\} \hspace{1mm} | \hspace{1mm} m\neq n$}, \label{prob_1c}
\end{align}
\end{subequations}
where the scalars $\nu_k$ are the optimization weights that the attacker can exploit to adjust the levels of interference to the different users. Note that constraint \eqref{prob_1b} sets the amplitudes of the reflection coefficients to one, modeling the passive operation of the RIS, while \eqref{prob_1c} ensures that $\bm{\Theta}$ is a diagonal matrix. These constraints, plus the matrix form of the objective function in \eqref{prob_1a}, make the problem difficult to solve. To tackle this complication, the attacker transforms the original formulation by exploiting the Khatri-Rao factorization:
\begin{align}\label{kt_id}
\left(\mathbf{B}^T \diamond \mathbf{A} \right) \mathrm{vecd}\{\mathbf{X}\} = \mathrm{vec}\{ \mathbf{A}\mathbf{X}\mathbf{B}\},     
\end{align}
where $\mathbf{X}$ is diagonal and $\mathbf{A}$ and $\mathbf{B}$ are arbitrary matrices of compatible dimensions. Specifically, by invoking \eqref{kt_id}, the following transformations can be applied:
\begin{align}
  \bm{\theta} &\triangleq \mathrm{vecd}\{\bm{\Theta}\} \in \mathbb{C}^{L},&
   \mathbf{S}_{k} \triangleq \mathbf{G}^T \diamond \mathbf{F}_{k}^H \in \mathbb{C}^{M N \times L}. \nonumber
\end{align}
Then, by plugging the above definitions into \eqref{prob_1}, the attacker achieves the following equivalent problem
\begin{subequations}\label{prob_2}
\begin{align}
    & \underset{\bm{\theta}}{\max} 
     \sum_{k=1}^{K} \nu_{k} \left\| \mathbf{S}_{k}\bm{\theta}
    \right\|_2^2,
    \label{prob_2a}\\[-1mm]
    &\text{s.t.}  \hspace{1mm} |[\bm{\theta}]_{n}| = 1, \hspace{1mm} \forall n \in\{ 1,\cdots, L\}.\label{prob_2b}
\end{align}
\end{subequations}
To simplify further the above optimization, the weighted sum in \eqref{prob_2a} is transformed into a matrix equivalent form by vertically concatenating each term of the sum, as follows
\begin{subequations}\label{prob_3}
\begin{align}
    & \underset{\bm{\theta}}{\max} 
     \left\| \begin{bmatrix} \sqrt{\nu_{1}} \mathbf{S}_{1} \\[-2mm]
     \vdots \\[-1mm]
     \sqrt{\nu_{K}} \mathbf{S}_{K}
    \end{bmatrix} \bm{\theta} \right\|_2^2,
    \label{prob_3a}\\[-1mm]
    &\text{s.t.}  \hspace{1mm} |[\bm{\theta}]_{n}| = 1, \hspace{1mm} \forall n \in\{ 1,\cdots, L\}.\label{prob_3b}
\end{align}
\end{subequations}
The problem in \eqref{prob_3} is non-convex due to the element-wise modulus constraint and is, in general, NP-hard \cite{Tranter17}. To overcome the challenge, we propose a suboptimal but practical strategy utilizing a projected gradient-based algorithm. The proposed strategy is presented in Algorithm 1, where $T$ is the number of iterations, $\beta$ is a parameter used to tune the step size $\alpha$, and $\lambda_{\text{max}}(\cdot)$ returns the largest eigenvalue of its argument. As demonstrated in \cite[Theorem 1]{Tranter17}, this algorithm converges to a Karush-Kuhn-Tucker (KKT) point of the original problem as long as $\alpha<1/\lambda^*$, with $\lambda^*$ denoting the largest eigenvalue of $\mathbf{\bar{S}}^H\mathbf{\bar{S}}$, defined in line $1$ of Algorithm 1.

It should be noted that the solution to \eqref{prob_3} represents a feasible attack that may cause significant performance degradation to the users, as demonstrated in our simulation results. However, it does not represent a system-wide optimal RIS attack, for it does not assume that the attacker has access to important parameters, such as BS precoders and BS-U channel matrices, which may be difficult, if not impossible, to obtain by the attacker.

\begin{figure}
\centering
\begingroup
\csname @twocolumnfalse\endcsname
\resizebox{.45\textwidth}{!}{%
\begin{minipage}{.56\textwidth}
\setlength{\algomargin}{5mm}
\setlength{\interspacetitleboxruled}{1mm}
\begin{algorithm}[H]
	\SetKwRepeat{Do}{do}{while}
	\SetAlgoLined
	\KwIn{\small $T, \beta \in (0,1), \{\nu_1, \cdots, \nu_K\}, \{\mathbf{S}_1, \cdots, \mathbf{S}_K\}$\;\vspace{1mm}}

	\small
	Initialize: $ 
  \mathbf{\bar{S}} = \begin{bmatrix} \sqrt{\nu_{1}} \mathbf{S}_{1} \\[-2mm]
     \vdots \\[-1mm]
     \sqrt{\nu_{K}} \mathbf{S}_{K}
    \end{bmatrix}$, $\bm{\theta}_{(1)} = \mathbf{I}_{L,1}$, $\lambda^{\star} = \lambda_{\text{max}}\left(\mathbf{\bar{S}}^H \mathbf{\bar{S}}\right)$, $\alpha = \frac{\beta}{\lambda^{\star}}$\;\vspace{1mm}
        
		\For{$t = 1, 2, \cdots, T - 1$}{\Indmm 
		Compute the gradient of \eqref{prob_3a}: \hspace{50mm}
		$ \bm{\delta}_{(t)} = \mathbf{\bar{S}}^H \left(\mathbf{\bar{S}} \bm{\theta}_{(t)} \right)$\;\vspace{1mm}
		
		Perform the gradient step: \hspace{50mm}
        $\bm{\phi}_{(t+1)} = \bm{\theta}_{(t)} + \alpha \bm{\delta}_{(t)} $\;\vspace{1mm}
				
		Compute the projection onto the unit 1-sphere:\hspace{50mm}
        $\bm{\theta}_{(t+1)} = e^{j\angle\left( \bm{\phi}_{(t+1)} \right)}$\;
		
		}\vspace{2mm}\KwOut{\small$\bm{\Theta} = \mathrm{diag}\{\bm{\theta}_{(T)}\}$.}
  
	\caption{\hspace{-1mm} Malicious RIS optimization for multi-user \\jamming attacks in mMIMO systems}\label{alg1}
\end{algorithm}
\end{minipage}
}%
\endgroup 
\end{figure}

\vspace{-1mm}
\section{Reception and Attack Mitigation}\label{sec_rec}
The BS precoders alone cannot cancel the attack by the RIS. This section addresses this issue by proposing two detection strategies to be deployed on the user side. In particular, we propose a two-layer reception strategy of the form $\mathbf{Q}_k\mathbf{W}_k$, where $\mathbf{W}_k$ is responsible for tackling the malicious RIS reflections and $\mathbf{Q}_k$ is a reception matrix responsible for removing inter-symbol interference. Moreover, while the \ac{CSI} of the link BS-U is available to the network, we assume that only an estimate of the cascade BS-RIS-U channel $\mathbf{Z}_{k}^H \triangleq \mathbf{F}_{k}^H \bm{\Theta}\mathbf{G}$ can be made available to the users\footnote{Advanced interference estimation techniques can be employed in the users' devices to estimate the cascade RIS channels. However, this goes beyond the objectives of this work and arises as an interesting research direction.}, which we model as follows
\begin{align}\label{eq_imp_csi}
    \mathbf{\hat{Z}}_{k} \triangleq \sqrt{1 - \tau^2} \mathbf{Z}_{k} + \tau \mathbf{E}_{k},
\end{align}
where $\mathbf{E}_{k}$ is the error matrix, which is independent of $\mathbf{Z}_{k}$ and whose entries follow the standard complex Gaussian distribution, and $\tau \in [0,1]$ models the level of error between the estimate $\mathbf{\hat{Z}}_{k}$ and the true channel $\mathbf{Z}_{k}$.\vspace{-3mm}

\begin{figure*}[t]
	\centering
	\includegraphics[width=.85\linewidth]{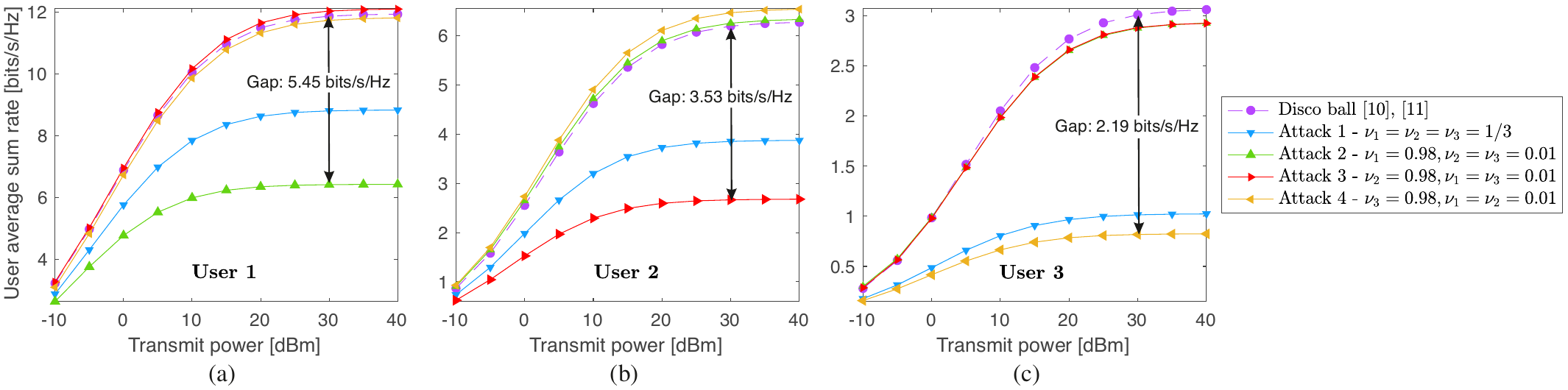}
	\caption{User average sum rate under different attack scenarios: (a) for user 1, (b) for user 2, and (c) for user 3, considering $N=4$ and $L = 200$.}\label{user_srates}\vspace{-4mm}
\end{figure*}

\subsection{\textsc{F}ull \textsc{MIT}igation (F-MIT) of RIS attacks}
By considering all RIS-reflected signals as interference, this first strategy aims to fully mitigate the attack. To this end, let us define $\mathbf{\Tilde{P}} \triangleq [\mathbf{P}_{1} \hspace{1mm} \cdots \hspace{1mm} \mathbf{P}_{K}] \in \mathbb{C}^{M \times K S}$ and $\mathbf{\Tilde{x}} \triangleq [\mathbf{x}_{1}^T \hspace{1mm} \cdots \hspace{1mm} \mathbf{x}_{K}^T]^T \in \mathbb{C}^{K S}$. Then, \eqref{rec_sig_02} can be rewritten as
\begin{align}\label{rec_sig_03}
\mathbf{y}_{k} = \mathbf{H}_{k}^H \mathbf{P}_{k}\sqrt{P}\mathbf{x}_{k} + \mathbf{Z}_{k}^H
     \mathbf{\Tilde{P}}\sqrt{P}\mathbf{\Tilde{x}} + \mathbf{n}_{k}.
\end{align}

Now, to mitigate the attack, the $k$th user needs an inner receptor $\mathbf{W}_k$ that can achieve $\mathbf{W}_k \mathbf{Z}_{k}^H \mathbf{\Tilde{P}} \approx \mathbf{0}_{S, S}$. Thus, $\mathbf{W}_k$ should lie in the left null space of the transformed channel matrix $\underline{\mathbf{Z}}_k \triangleq \mathbf{Z}_{k}^H \mathbf{\Tilde{P}} \in \mathbb{C}^{N \times K S}$. Since the perfect knowledge of $\mathbf{Z}_{k}$ is not possible, users approximate the desired left null space from the left eigenvector matrix $\mathbf{U}_{\underline{\mathbf{\hat{Z}}}_k}  \in \mathbb{C}^{N\times N}$ of the estimated channel $\underline{\mathbf{\hat{Z}}}_k \triangleq \mathbf{\hat{Z}}_k^H \mathbf{\Tilde{P}}  \in \mathbb{C}^{N \times K S}$. Note that, in addition to canceling interference, the matrix $\mathbf{W}_k$, with dimension $W \times N$, should satisfy $N \geq W \geq S$ for $\mathbf{Q}_k$ to remove the inter-symbol interference. Provided that $N > KS$ is satisfied, $\underline{\mathbf{\hat{Z}}}_k$ will have full column rank and a left null space of dimension $N - KS$. As a result, the perfect orthogonality with the estimated channel, i.e., $\mathbf{W}_k \underline{\mathbf{\hat{Z}}}_k = \mathbf{0}_{W, S}$, is achieved if and only if $W \leq N - KS$, otherwise residual interference is left. With these observations, $\mathbf{W}_k$ can be constructed from the last $W$ rows of $\mathbf{U}^H_{\underline{\mathbf{\hat{Z}}}_k}$, as follows
\begin{align}\label{prec_str1}
\mathbf{W}_k = \left[\mathbf{U}^H_{\underline{\mathbf{\hat{Z}}}_k}  \right]_{\left(N - W + 1\right) : N, :} \in \mathbb{C}^{W \times N},    
\end{align}
where we should satisfy $N > KS$ for the existence of a non-trivial left null space. Moreover, we adjust $W = \max\{N - KS, S\}$, which satisfies the requirement $W \geq S$.

Now, since the resulting effective channel $\mathbf{W}_k\mathbf{H}_{k}^H \mathbf{P}_{k}$ is an $W \times S$ matrix, with $S \leq W$, we can remove the inter-symbol interference by computing its left inverse, as follows
\begin{align}\label{posrec_str1}
\mathbf{Q}_{k} =\hspace{-.5mm} [(\mathbf{W}_k\mathbf{H}_{k}^H \mathbf{P}_{k})^H \mathbf{W}_k\mathbf{H}_{k}^H \mathbf{P}_{k}]^{-1}(\mathbf{W}_k\mathbf{H}_{k}^H \mathbf{P}_{k})^H. \hspace{-.8mm}
\end{align}

Then, after filtering the signal in \eqref{rec_sig_03} with \eqref{prec_str1} and \eqref{posrec_str1}, the $k$th user detects the $s$th symbol with the following SINR
\begin{align}
     \gamma_k^s &= \frac{P |[\mathbf{Q}_k\mathbf{W}_k\mathbf{H}_{k}^H \mathbf{P}_{k}\mathbf{x}_{k}]_s|^2}{ P|[\mathbf{Q}_k\mathbf{W}_k\mathbf{Z}_{k}^H
      \mathbf{\Tilde{P}}\mathbf{\Tilde{x}}]_s|^2  + \sigma^2 [\mathbf{Q}_k \mathbf{Q}_k^H]_{ss}},
\end{align}
where the rightmost term in the denominator follows from $[\mathbf{Q}_k\mathbf{W}_k\mathbf{n}_{k}]_s|^2 = \sigma^2 [\mathbf{Q}_k\mathbf{W}_k\mathbf{W}_k^H \mathbf{Q}_k^H]_{ss} = \sigma^2 [\mathbf{Q}_k \mathbf{Q}_k^H]_{ss}$, given that $\mathbf{W}_k$ is a semi-unitary matrix, i.e., $\mathbf{W}_k\mathbf{W}^H_k = \mathbf{I}_{S}$.\vspace{-1mm}

\subsection{\textsc{H}arnessing and \textsc{MIT}igation (H-MIT) of RIS attacks}

In this subsection, we propose a second approach that relaxes the orthogonality constraint $W \leq N - SK$ that is demanded by the F-MIT strategy. In this method, only unintended reflected streams are selected for mitigation. As an added advantage, users can harness the RIS reflections to extract their own symbols. To attain this objective, we define $\mathbf{\Tilde{P}}^{\star} \triangleq [\mathbf{P}_{1} \hspace{1mm} \cdots \hspace{1mm} \mathbf{P}_{k-1} \hspace{1mm} \mathbf{P}_{k+1} \hspace{1mm} \cdots \hspace{1mm} \mathbf{P}_{K}] \in \mathbb{C}^{M \times (K-1) S}$, and $\mathbf{\Tilde{x}}^{\star} \triangleq [\mathbf{x}_{1}^T \hspace{1mm} \cdots \hspace{1mm} \mathbf{x}_{k-1}^T \hspace{1mm} \mathbf{x}_{k+1}^T \cdots \hspace{1mm} \mathbf{x}_{K}^T]^T \in \mathbb{C}^{(K-1) S}$, which allow us to rewrite the signal in \eqref{rec_sig_03} as
\begin{align}\label{rec_sig_05}
\mathbf{y}_{k} = (\mathbf{H}_{k}^H + \mathbf{Z}_{k}^H) \mathbf{P}_{k}\sqrt{P}\mathbf{x}_{k} + \mathbf{Z}_{k}^H \mathbf{\Tilde{P}}^{\star}\sqrt{P}\mathbf{\Tilde{x}}^{\star} + \mathbf{n}_{k}.
\end{align}

Thus, we need to achieve $\mathbf{W}_k \mathbf{Z}_{k}^H \mathbf{\Tilde{P}}^{\star} \approx \mathbf{0}_{W, S}, \forall k \in \mathcal{K}$, which can be accomplished using a similar approach of that employed for the F-MIT strategy. Specifically, let $\underline{\mathbf{\hat{Z}}}^{\star}_k \triangleq \mathbf{\hat{Z}}_k^H \mathbf{\Tilde{P}}^{\star}  \in \mathbb{C}^{N \times (K-1) S}$ be the estimated transformed interference channel, and denote by $\mathbf{U}_{\underline{\mathbf{\hat{Z}}}^{\star}_k}$ the associated left eigenvector matrix obtained from the SVD of $\underline{\mathbf{\hat{Z}}}^{\star}_k$. Then, $\mathbf{W}_k$ can be obtained from the last $W$ rows of $\mathbf{U}^H_{\underline{\mathbf{\hat{Z}}}^{\star}_k}$, as follows
\begin{align}\label{prec_str2}
\mathbf{W}_k = \left[\mathbf{U}^H_{\underline{\mathbf{\hat{Z}}}^{\star}_k}  \right]_{\left(N - W + 1\right) : N, :} \in \mathbb{C}^{S \times N},
\end{align}
with $W = \max\{N - (K-1)S, S\}$, where now we need $N > (K-1) S$ for achieving the desired non-trivial left null space, and only $W \leq N - (K-1) S$ for the orthogonality with the estimated selected interference channels.

By relying on the effectiveness of $\mathbf{W}_k$ to tackle $\mathbf{Z}_{k}^H \mathbf{\Tilde{P}}^{\star}\sqrt{P}\mathbf{\Tilde{x}}$ in \eqref{rec_sig_05}, users address the remaining inter-symbol interference with the left inverse of the resulting estimated projected channel $\mathbf{W}_k(\mathbf{H}_{k}^H + \mathbf{\hat{Z}}_{k}^H)\mathbf{P}_{k} \in \mathbb{C}^{W\times S}$, which is given by
\begin{align}\label{posrec_str2}
\mathbf{Q}_{k} &= [(\mathbf{W}_k(\mathbf{H}_{k}^H + \mathbf{\hat{Z}}_{k}^H)\mathbf{P}_{k})^H \mathbf{W}_k(\mathbf{H}_{k}^H + \mathbf{\hat{Z}}_{k}^H)\mathbf{P}_{k}]^{-1}\nonumber\\
&\times (\mathbf{W}_k(\mathbf{H}_{k}^H + \mathbf{\hat{Z}}_{k}^H)\mathbf{P}_{k})^H.
\end{align}

Note in \eqref{posrec_str2} that the design of $\mathbf{Q}_{k}$ now depends on $\mathbf{\hat{Z}}_{k}$. This implies that users might also experience inter-symbol interference. By applying \eqref{posrec_str2} to the signal in \eqref{rec_sig_05}, the $k$th user recovers the $s$th symbol with the following SINR
\begin{align}
     \gamma^{s}_{k}
     & = \frac{P |[\bm{\Xi}_k \mathbf{x}_{k}]_s|^2}{P |[\bm{\Upsilon}_k \mathbf{x}_{k}]_s|^2 \hspace{-1mm} + \hspace{-.5mm} P|[\mathbf{Q}_k\mathbf{W}_k\mathbf{Z}_{k}^H
     \mathbf{\Tilde{P}}^{\star}\mathbf{\Tilde{x}}^{\star}]_s|^2 \hspace{-1mm} + \hspace{-.5mm} \sigma^2 [\mathbf{Q}_k \mathbf{Q}_k^H]_{ss}}.
\end{align}
where $\bm{\Xi}_k \triangleq \mathrm{diagm}\{\mathbf{Q}_k\mathbf{W}_k(\mathbf{H}_{k}^H + \mathbf{Z}_{k}^H) \mathbf{P}_{k}\}$ is a diagonal matrix with the effective channel coefficients for intended symbols, and $\bm{\Upsilon}_k \triangleq \mathbf{Q}_k\mathbf{W}_k(\mathbf{H}_{k}^H + \mathbf{Z}_{k}^H) \mathbf{P}_{k} - \bm{\Xi}_k$ is a hollow matrix with the coefficients for inter-symbol interference.

\section{Simulation Results}
In this section, we report on the outcomes of Monte Carlo simulations. First, we compare the user sum rates of a \ac{mMIMO} system exposed to RIS attacks leveraging random phase shifts, a.k.a. Disco Ball attacks \cite{huang2023illegal,huang2023disco}, and a system under attacks launched by our malicious RIS scheme considering four different threat scenarios. Subsequently, we compare the performance of a safe mMIMO (free from threats) and a system under our optimized attack with and without employing the proposed mitigation strategies, F-MIT and H-MIT.

\begin{figure*}[t]
	\centering
	\includegraphics[width=1\linewidth]{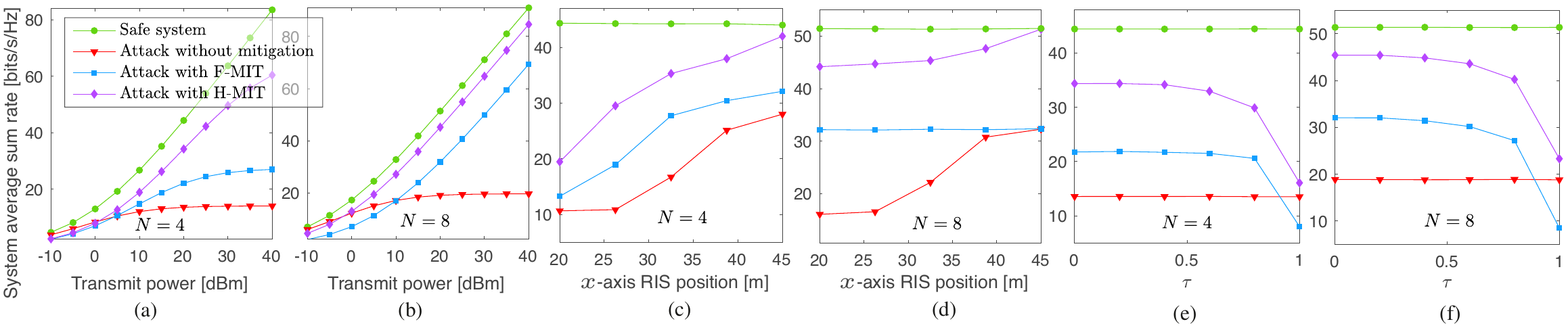}
	\caption{System average sum rate (with $L=200$ and $\nu_1 = \nu_2 = \nu_3 = 1/3$): (a) for $N=4$ and (b) for $N=8$ when $\tau = 0$; (c) for $N=4$ and (d) for $N=8$, considering different $x$-axis RIS positions and a fixed $y$-axis coordinate of $20$~m, with $P = 20$~dBm and $\tau = 0$; (e) for $N=4$ and (f) for $N=8$ with different values of $\tau$ and $P = 20$~dBm. }\label{system_srates}\vspace{-5mm}
\end{figure*}

In each case, the BS is equipped with a uniform linear array of $M = 60$ antennas, and it communicates with $K = 3$ multi-antenna users, such that $S = 2$ parallel data symbols are transmitted for each user. Moreover, we assume that the BS is located at the origin, whereas users $1, 2$, and $3$ are located at the coordinates $(20, 0)$~m, $(20, 40)$~m, and $(50, 20)$~m, respectively. As for the attacker, unless otherwise stated, its malicious RIS is strategically deployed at the coordinate $(30, 20)$~m. With this geometrical scenario, the path-loss coefficients for the links BS-RIS, RIS-U, and BS-U are, respectively, computed as $(d^{\text{\tiny BS-RIS}})^{-\eta}$, $(d^{\text{\tiny RIS-U}}_{k})^{-\eta}$, and $(d^{\text{\tiny BS-U}}_{k})^{-\eta}$, for $k\in \{1,2,3\}$, where $d^{\text{\tiny BS-RIS}}$, $d_k^{\text{\tiny RIS-U}}$, and $d_k^{\text{\tiny BS-U}}$ are the distances, and $\eta$ is the path-loss exponent set to $2.5$ for all links. Regarding Algorithm \ref{alg1}, we set the step parameter to $\beta = 0.99$ and the number of iterations to $T = 3\times 10^3$. Furthermore, we adjust the noise variance to $\sigma^2 = -40$~dBm and employ a uniform power allocation among users and symbols, such that $\alpha_{k,1} = \cdots = \alpha_{k, S} = 1$, $\forall k \in \{1,2,3\}$. 

Fig. \ref{user_srates} presents the user average sum rate curves, computed as $R_k = \mathrm{E}\{\sum^S_{s=1} \log_2 (1 + \gamma_k^s)\}$, which informs the sum spectral efficiency for each user $k \in \{1,2,3\}$ under different attack scenarios. These results validate the proposed malicious RIS algorithm and demonstrate its efficiency and adaptability in executing targeted attacks against a specific user. As can be seen, this capability cannot be achieved with the Disco Ball attack. For instance, in Fig. \ref{user_srates}(a), user 1 can achieve a sum rate of $8.79$~bits/s/Hz when the transmit power is $30$~dBm under the proposed RIS attack with equal weights (Attack 1). When the attacker targets this user, i.e., $\nu_1 = 0.98$ in Attack 2, its sum rate drops to only $6.41$~bits/s/Hz, a $2.38$~bits/s/Hz degradation compared to the equal-weighted case and impressive $5.45$~bits/s/Hz compared to the Disco Ball attack. Similar behavior can also be verified for the other users in Figs. \ref{user_srates}(b) and \ref{user_srates}(c), where the targeted attacks launched by our optimized scheme are remarkably more powerful than the baseline counterpart.

Fig. \ref{system_srates} evaluates the system average sum rate, calculated as $\bar{R} =\mathrm{E}\{\sum^K_{k=1} \sum^S_{s=1} \log_2 (1 + \gamma_k^s)\}$. Specifically, Figs. \ref{system_srates}(a), \ref{system_srates}(c), and \ref{system_srates}(e) show that even when the constraints of \eqref{prec_str1} and \eqref{prec_str2} cannot be met, i.e., when $N=4$, both F-MIT and H-MIT can still provide significant performance gains over the case without mitigation, as long as the transmit power is high enough. When $N=8$, the perfect orthogonality with the estimated RIS channel can be achieved by both F-MIT and H-MIT, removing the saturation on the sum rate curves, as shown in Fig. \ref{system_srates}(b). However, since these receptors also amplify the noise, attaining the same performance as the safe system is impossible. Nonetheless, we can see that recycling intended signals from the RIS-reflected signals is advantageous as it enables the H-MIT receptor to significantly outperform the F-MIT counterpart for $N=4$ and $N=8$. These performance gains are also reproduced in the subsequent figures. In Figs. \ref{system_srates}(c) and \ref{system_srates}(d), the RIS is moved to different positions along the $x$-axis direction, while keeping its $y$-axis coordinate fixed at $20$~m. As can be seen, the closer the RIS is to the BS, the stronger the performance degradation caused by the proposed attack. Last, Figs. \ref{system_srates}(e) and \ref{system_srates}(f), present the system sum rates for the case when the estimation of the RIS channels is imperfect, revealing that the H-MIT receptor is beneficial even when the estimation error is high.

\section{Conclusions}
In this letter, we have shed light on a powerful adversarial attack that can be launched with an optimized RIS against multiple users without requiring their legitimate channels. %
Simulation results demonstrated the adaptability and efficiency of the proposed RIS algorithm in significantly degrading the sum rates of users, highlighting its superiority over the state-of-the-art Disco Ball attack. This stresses the significance of developing countermeasures to safeguard next-generation wireless systems from emerging physical layer threats. In this regard, we proposed two defense mechanisms, F-MIT and H-MIT, for countering such malicious RIS attacks. Both receptors provided significant performance enhancements over unmitigated attack scenarios, even when their null space constraints could not be satisfied. Still, it is worth noting that users in low transmit power regimes may still be affected by the attack, and additional research and appropriate mitigation strategies are required to safeguard them. Our future work shall delve deeper into estimation strategies for enabling this and related RIS attacks and developing new robust mitigation strategies to counteract threats under realistic CSI conditions.





\ifCLASSOPTIONcaptionsoff
\newpage
\fi

\bibliographystyle{IEEEtran}
\bibliography{bibtex/references}

\end{document}